# Emergent Probability
A Directed Scale-Free Network Approach to Lonergan's Generic Model of Development

**Michael Bretz**
Department of Physics
University of Michigan
mbretz@umich.edu

## 1.1  Introduction

An intriguing heuristic model of development, decline, and change conceived by Bernard J.F. Lonergan (BL) in the late 1940's was laid out in a manner now recognizable as representing an early model of complexity. This report is a first effort toward eventually translating that qualitative vision, designated *Emergent Probability*\* (EP), into a viable network computer study.

In his study of human understanding, Lonergan [1992] saw the task of constructing a cohesive body of explanatory knowledge as a convoluted building process of *recurrent schemes* (RS) that act as foundational elements to further growth.  Although BL's kernal RS was composed of the cognitional dynamics surrounding *Insight* [Bretz, 2002], other examples of recurrent growth schemes abound in nature: resource cycles, motor skills, biological routines, autocatalytic processes, etc. The corresponding growing generic *World Process* can alternatively be thought of as chemical, environmental, evolutionary, social, organizational, economical, psychological, or ethical [Melchin, 2001], and its generality might be of particular interest to complex systems researchers.

*Schemes of Recurrence* are conjoined dynamic activities where, in simplest form, each element generates the next action, which in turn generates the next, until the last dynamic regenerates the first one again, locking the whole scheme into long term stable equilibrium. BL modeled generic growth as the successive appearance of *conditioned Recurrent Schemes* (RS), each of which comes into function with high probabilistically once all required prior schemes have become functional. RS's can be treated as dynamic black cells of activity which themselves may contain internal structures and dynamic schemes of arbitrary complexity.  Emergent Probability is a generic heuristic model.  Applications to specific physical problems requires detailed knowledge of the recurrent schemes' makeup and of their interrelationships (an elementary example is provided in the Appendix).

The universe of possibilities available to any open-ended development process is vast, and the interrelations among the elements can become arbitrarily convoluted as the process proceeds. All growth in that universe, no matter how complex, depends upon a suitable underlying environmental "situation", and an ecology, or niche, in order to thrive. Any full simulation of EP must allow for adaptation to a changing situation and ecology, so that recovery can occur when events disrupt underlying schemes, or when separate growths vie among themselves for dominance within a stressed environment

Here, I present first results obtained from an exploratory toy model of EP development. This MATLAB simulation uses a scale-free, directed network (nodes as RS's, inward links from the conditional RS's) to represent the universe of relational possibilities. Growth occurs when nodes of the network are sequentially activated to functionality. The appearance of RS clusters (*Things*), their dependence on the underlying ecology and situation, and their interplay are sought.

## 1.2 Modeling the Potential World

A sparse adjacency matrix was grown that defines all of the possible nodes of the toy model simulation, their link dependencies on selected previously grown nodes and their role in constraining younger nodes. Each row/column number names a specific, unique node and values of 1 in the sparse matrix elements represent individual links between adjacent nodes of the network. The network was grown from 3 core nodes by combining aspects of the scale-free growing network methods of [Dorogovtsey, 2000] and [Krapivsky, 2001]. For each succeeding iteration, with probability $q = .83$, a new node without links was created having an attractiveness $A = 1$. With the complementary probability, $p = .17$, $m = 3$ new links were added between statistically chosen existing nodes. Selection of the m target nodes and of originating nodes used weightings directly proportional to $(k_{in} + A)$ and $(k_{out} + A)$, where the k's refer to the number of inward and outward links associated with individual nodes, respectively.

The resulting square matrix containing a total of 700,000 undirected nodes was altered to ensure that all of the links pointed toward younger nodes. This was accomplished by superimposing the transpose of the upper triangular portion of the matrix onto its lower triangular portion before isolated nodes and nodes having zero or one in-link were winnowed from the matrix. After compacting, the final triangular adjacency matrix representing the universe of possibilities for our toy model contained 99,059 nodes and 277,493 directed links, with zeros along the diagonal. Although not huge, the adjacency matrix, named WORLD, is still large enough that elements of the RS growth activity can be explored before being unduly quenched by finite size limitations
.

## 1.3 Characterizing the WORLD Matrix

The node link distributions, P(k), representing the total number of nodes having specific $k_{in}$ and $k_{out}$ are presented in a log-log graph shown in Fig. 1. Both distributions follow power law behavior from $k \cong 4$ with exponents $\gamma \sim 1.64$ and $\gamma \sim 8$ for the out-link and in-link distributions, respectively. Significantly, the oldest two nodes possesses an appreciably larger number of outlinks than is consistent with the out-degree scale-free trend. These are interpreted as defining the 'situation' under which the recurrent scheme nodes will grow upon WORLD. Traversing all directed link paths emanating from node 1 reveals a degree of separation, d, of 5 steps to all existing nodes in WORLD. Following the link paths from node 6 gives d = 7 steps. Probing other link paths indicates a maximum of $d \cong 18$ directed steps for the WORLD matrix.

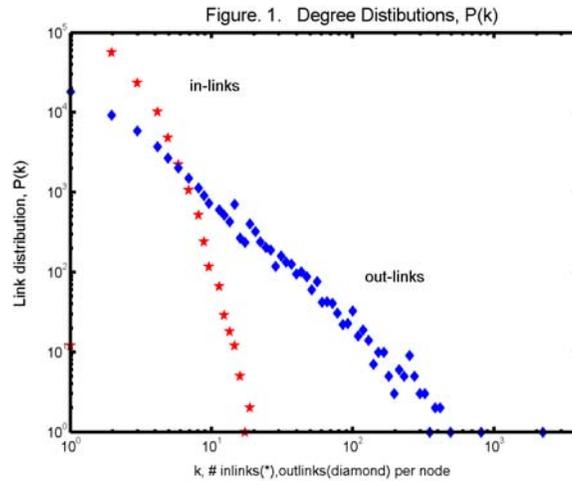

Another parameter that characterizes networks is the correlation coefficient, CC = (# links between parent nodes)/(# possible links), which measures the local connectivity between constraining in-link (or out-link) nodes. A coefficient near 1 would indicate a highly connected network. The WORLD matrix, however, possesses very low correlation coefficients for parent nodes and also for child nodes, with values $CC_{in}$ = .0028, $CC_{out}$ = 7.0 x $10^{-6}$, respectively.

## 2.1 Development as World Process

Our *World Process* in *Emergent Probability* is the probabilistic activation to functionality of nodes upon the WORLD matrix as the situation, ecology (to be defined), and node conditioning permit. Recurrent Schemes become *virtually unconditioned* toward activation when all of their *conditions* have been satisfied by the actual functioning of their originating in-link nodes. Starting from the three *formally unconditioned* WORLD nodes, #'s 1, 2 and 6 that possess no in-links, each iteration of EP visits and interrogates all non-functioning nodes.

Activation of RS's to functionality is probabilistic. All RS's have some small probability, $p_i$, of irregularly receiving actions from their non-functioning constraints. The probability, P, for all k actions to occur simultaneously and bring that particular RS to functionality is

$$P = \prod_i (p_i). \qquad (1)$$

Should that RS instead become virtually unconditioned, the probability for functionality of its scheme suddenly jumps to a much larger value, for <u>if any one</u> of the several conditional node actions starts the scheme, then the scheme will emerge to functional stability. P can then be better written as one minus the product of probabilities for conditional nodes <u>not</u> initiating action (see Appendix),

$$P = 1 - \prod_i (1-p_i) \sim \sum_i (p_i). \qquad (2)$$

On other words, once the parent schemes are functioning, the probability for a scheme's appearance leaps from the product of the separate action probabilities of its parents to their sum, making the chances of its appearance hugely more likely!

For calculational purposes, each node possesses an individual action probability, $p_i$, which was assigned randomly within a Rayleigh distribution envelope having mean value = 0.1 and a long low tail to higher values (just choosing the probabilities as random numbers between 0 and some small number < 1 seems quite unphysical). Unlike Lonergan's EP, the assumptions made for this study are that i) the P's from eq. (1) are all vanishingly small, ii) only a single RS of each type can be functional, but iii) all individual RS's may share their dynamics with many dependent RS's at the same time

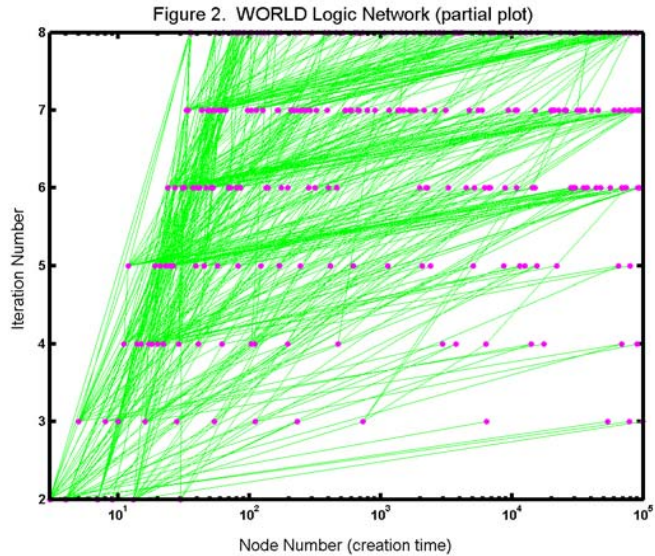

Figure 2. WORLD Logic Network (partial plot)

## 2.2 Characterizing the Growing Node Network

It is instructive to initially grow RS's upon the WORLD matrix with all appearance probabilities set to 1. This corresponds to a logic matrix where all activations to function that might occur at every iteration, will actually occur. Figure 2 is a map of such a logical run. The respective functional node numbers lie along the log(node #) axis and the iteration # is along the y-axis. Thus, every potential node of WORLD can be classified by its logical iteration number. Also shown in the figure are all connecting links to nodes activated during the first 8 iterations. (Notice that all links point toward higher numbered (younger) nodes and higher iteration #s.) Since the links already dominate the scene after several iterations, showing the entire plot with associated links would not be useful. (The full plot occupies a triangular area extending to iteration 47 and node numbers $\sim 10^5$ at which point all nodes are finally functional.) The effective EP network diameter, then, is much larger than the WORLD matrix's maximum degree of separation ($d \cong 18$) that was determined by just jumping along directed nodes .

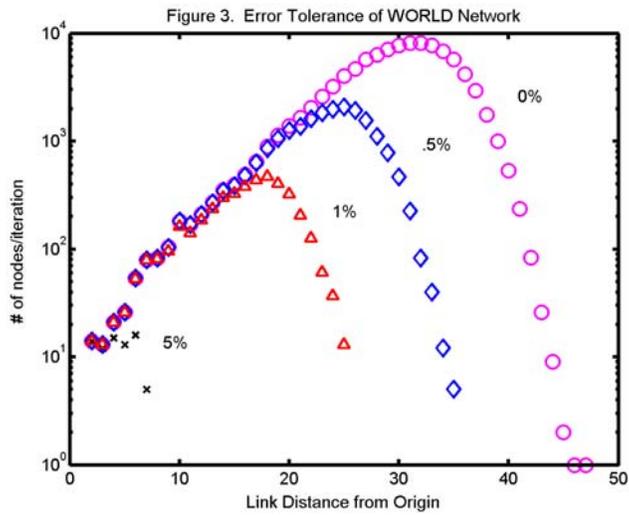

A second instructive procedure is to randomly specify a small number of nodes as permanently non-functional. These nodes strongly poison and fragment the whole growth process, as is expected for scale-free networks [Albert, 2000]. Activation growth rates under these logic conditions are shown in fig. 3 for four impurity concentrations (0%, 0.5%, 1% and 5%). For the pure case all nodes are eventually accessed by iteration 47, but even slight impurities appreciably cut the accessibility to nodes.

Now we shall activate nodes probabilistically (rather than logically), starting with the formally unconditioned nodes 1, 2 and 6. With each iteration more and more nodes spring to functionality (as recorded by 1's along the sparse WORLD diagonal), as they are driven by an ever increasing state space of possibilities. The log of this simple nodal function growth rate is shown as the black line in Fig. 4. The rate rises as a steep exponential for the first 55 or so iterations (to 2004 total functioning nodes), before settling in to a lesser, but still substantial exponential growth rate. But then, after another 55 iterations and some 44,000 functioning nodes later there appears a rounding off (and eventual descent, not shown). The ever-decreasing state space density of the WORLD matrix starts to come into play from this point forward in the simulation. Results up to iteration 55 are kept as the starting point for subsequent runs, since initial growing pains seem to have subsided by then. One can think of the early iterations as a <u>baseline representing the world situation and ecology</u> upon which further activity occurs.

### 2.3  Toward Complexity

Although Lonergan envisioned the interplay of vast numbers of recurrent schemes operating across almost limitless regions of space and through enormous ranges in time, we are constrained here to a simple toy model without spatial dimensions and operating for only a few hundred time iterations. Here, we shall only perform a few simulations to observe an interplay between two equivalent growing clusters as a probe of Emergent Probability dynamics.

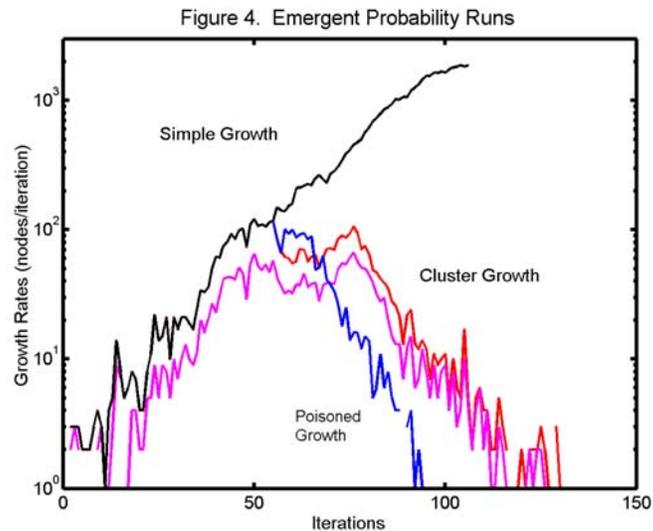

All ecology nodes are randomly allotted node strengths = +/-1 along the WORLD diagonal. The sign identity of each further node activation will be governed by a simple majority rule. That is, the activation of nodes are accompanied by sign designations determined by the sum of constraining nodes. Should that sum be zero for nodes having an even number of parents, then no sign decision is made and those sign-unresolved nodes are inactivated remain thereafter. The simulation (blue line in Fig. 4) demonstrates that the discarded nodes (923 in all) act as errors that poison the growing EP network, whose low error tolerance quenches the growth rate!

Simple competition between the two +/- signed clusters (*Things*) can only play out if the simulation contains dynamic mechanisms to thwart such quenching. A host of cleaver scenarios can be designed for the purpose, but presenting only one of them is sufficient to demonstrate a route toward increasing complexity for the system. The node identity problem is partially resolved by envisioning an ongoing process of merging two similar nodes into a single new signed node. When trying to activate a sign-unresolved node, a search is made among that node's parents to see if any of them have a pair (or more) of common out-links to a second, different node. If found, then with 20% probability, the in-links and out-links of the two nodes are merged, respectively, and copying the sign of the second node (if functional), is recorded in the WORLD matrix <u>overwriting</u> the unresolved node. The second node, if not yet functional, is barred from further participation.

Results are shown as the red line in Fig. 4. The growth rate recovered from an initial drop, but after another 10 iterations it collapsed, as if the node merging process itself limited the state space for growth. Also, the collapse is less steep than for the poisoned run (blue line). The two signed clusters grew at non-identical rates, the magenta line denoting the growth rate for the + cluster in Fig. 4. Until about iteration 75 its size remained somewhat more than half that of the total active nodes, but thereafter it grows in strength relative to the negative cluster (not shown), ending up as dominant.

Duel cluster growth can be further enhanced by assigning random strengths between +/-1 to each ecology node (those nodes functioning by iteration 55). Cluster designation takes place as before, with sign-unresolved nodes being those with $|\Sigma(\text{parent strengths})| <$ weakness, where this threshold is arbitrarily chosen as weakness = 0.05. Instead of inactivating below-threshold nodes, they are activated with randomly chosen strengths = (+/-)weakness (black dots). The resulting growth rate is shown as the red curve in Fig. 5. The rate now follows the simple exponential growth rate curve of Fig. 4 up to iteration 75 where there are 600 nodes/iteration, before dropping back exponentially. A total of 18840 nodes were activated to functionality by iteration 140. The growth rate for the + cluster is also displayed (magenta line in Fig. 5). Integrated growth totals rise exponentially for both clusters before leveling out to a cluster node ratio of 1.27, reflecting the bias from the 3 formally unconditioned situational nodes. The average strength/node for each iteration, a good measure of node fitness, becomes stable at 0.2+/-0.05, but after iteration 100 it gets very noisy. A repeat run allotting all unassigned nodes to the negative cluster eliminated the node ratio bias.

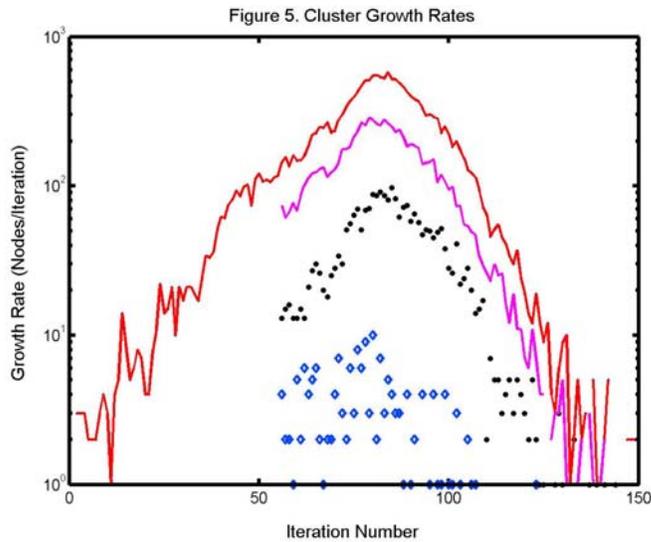

Figure 5. Cluster Growth Rates

The dynamics of cluster growth becomes clearer when the appearance rate of merged nodes is added to Fig.5 (diamonds). These new nodes apparently redirect future growth by strongly affecting the state space for development. In the present case when merging opportunities dry up, the cluster growth rate quenches, since it can no longer continue along the original simple trajectory of Fig 4.

Node merging (or alternate scenarios), then, will slowly evolve the highly sparse adjacency WORLD matrix, fundamentally altering the world logic network of Figure 2. The matrix will no longer remain fully triangular, but will record in-links to older nodes, creating closed directed adjacency matrix loops. A simple MATLAB eigen routine that seeks such pathways produced about as many degenerate eigenvalues as the number of merged nodes in our cluster simulation. The sudden appearance of these eigenvalues can be interpreted as the emergence of a higher order set of recurrent schemes that substantially increases world process complexity. The remerging of merged nodes, although not implemented in this network study, will act to *accelerate* complexity growth, enriching and deepening the ongoing Emergent Probability dynamics. A similar change is known to mark the emergence of autocatalytic sets in a recent complexity model of evolution [Jain, 1998].

## 3.1 Conclusion and Outlook

The heuristic model designated Emergent Probability introduced a half century ago by Bernard Lonergan is a powerful model of development, decline and change (aspects of which have evidently been reinvented as key elements in various present day approaches to complexity [Padgett, 1996] [Edelman,1987]). Although we have explored some of its basic underlying dynamics and glimpsed but one simple route toward complexity, the full EP model represents a genetic framework for the understanding of process. Moreover, the model greatly expands in reach when human, social and dialectical interactions are subsequently addressed. Emergent Probability methods, then, yield a vantage point for furthering the unification of disparate interdisciplinary fields and for effectively modeling the dynamics of human endeavors.

Work is continuing to enlarge the network simulation size, scope and efficiency. Key elements - growth disruption mechanisms, resource generation and competition, multiple niches, etc.- still need computer implementation. To properly simulate *Things* and their complex interactions also requires variable populations and the continual expansion of possibilities along with the ever-*higher integrations* of emergent complexity patterns. In future network models nodes of the WORLD matrix will be assigned inherent multiplicative growth fitness values, allowing late appearing nodes to bloom more fully [Bianconi, 2000]. Additionally, techniques for the tracking of EP's projected *upwardly but indeterminately directed dynamism* will be sought.

# References

The author wishes to thank D.W. Oyler and K.R. Melchin for conversations regarding Lonergan's Emergent Probability model.

# Appendix: Learning to ride a bike

As a simple example of how a nest of recurrent schemes may constrain, but sustain skill development, we schematically model the process of learning to ride a bicycle by a well-coordinated, inexperienced student (see diagram). Once learned, bike riding sustains a host of further activities.

The essential underlying *situation* is that: i) the student has good health, strength and motivation, ii) there is a suitable terrain and sufficiently good weather to ride, iii) an operable bicycle is available that is properly sized and adjusted, and iv) effective instruction is provided. The recurrent *functions* to be mastered are peddling, balancing, turning, braking and gear shifting. They allow their respective *actions* to recur over and over during bike rides.

First the instructor, Bernie, verbally lectures student, Mike, on how to perform each action. There is some probability, Bernie reasons, that he will start riding immediately, since each rider action is commonly known to first occur with finite probability, $p_i$. Let's assign $p_i$'s as peddling ~.75, balance ~ .1, turning ~ .3, braking ~.4, and shifting ~.4. Mike is given a push, tries to do all actions at once, and after a few wobbles, falls with a crash to the ground! That's because the combined probability of all required actions being performed correctly on the first try is miniscule ($P = \Pi\, p_i < .004$).

So, now Bernie reasons that in order to lit constraints to riding each of the five *actions* must be made to actually *function* through training the individual skills. He therefore has Mike thoroughly practice each separate activity by holding the bike seat and pushing him around. After a half hour of training, Mike again attempts to perform all actions at once without assistance after the initial push. This time he starts to wobble along on his own, quickly gaining stability and confidence.

Why has he succeeded on this second try? Because with all requisite skills now functional, Mike must successfully perform only ONE action, and all the rest will naturally fall into place. For instance, supposing Mike turned his handlebars a wee bit, this action could produce an increase in balance. The extra balance would allow him a chance to peddle some, which would increase his speed, which would complement his newly acquired balance skills, allowing other skills to also happen automatically. After a few moments he would be able to steer around obstacles and brake appropriately for safety as needed, completing the task. This cooperative function, viewed as a new emergent scheme of recurrence, is given the name "Biking".

Bernie is impressed and realizes that his calculations for the probability of Mike's success should be calculated quite differently now that the underlying skills have been mastered. Although Mike started out by trying to perform all of his new skills at once (having respective first occurrence probabilities as given above), one took hold and all other actions followed naturally. There is only a small probability of rider failure, $(1 - P)$, since any other action could have also initiated the task. Now, P equals one minus the product of all the separate failure probabilities, $(1 - p_i)$, or simplifying, $P \sim \Sigma\, p_i = .94$.

## Schematic Diagram of Bike Riding
**circles are recurrent schemes**

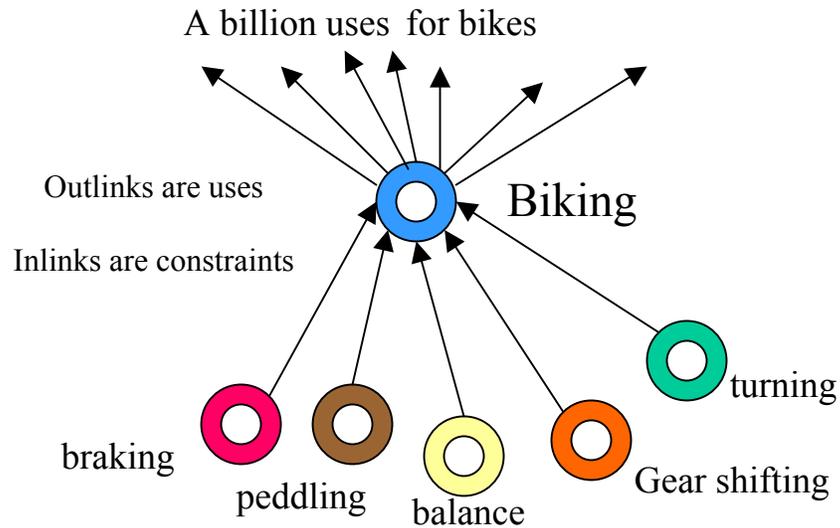